\documentclass[submission,copyright,creativecommons]{eptcs}
\providecommand{\event}{ThEdu'22} 

\usepackage{float}
\usepackage{xcolor}
\usepackage{qtree}
\usepackage{proof}
\usepackage{logicproof}
\usepackage{graphicx}
\usepackage{tikz}
\usepackage{pgf-pie}

\usepackage{caption}
\usepackage{subcaption}
\usepackage{iftex}

\ifpdf
  \usepackage{underscore}         
  \usepackage[T1]{fontenc}        
\else
  \usepackage{breakurl}           
\fi

\title{ANITA: Analytic Tableau Proof Assistant}
\author{Davi Romero Vasconcelos
\institute{UFC\\ Quixadá, Brazil}
\institute{School of Computer Science\\
Federal University of Ceará\\
Quixadá, Brazil}
\email{daviromero@ufc.br}
}
\def\titlerunning{ANITA: Analytic Tableau Proof Assistant}
\def\authorrunning{D. R. Vasconcelos}
\begin{document}
\maketitle

\begin{abstract}
This work presents the system ANITA (Analytic Tableau Proof Assistant) developed for teaching analytic tableaux to computer science students. The tool is written in Python and can be used as a desktop application, or in a web platform. This paper describes the logical system of the tool, explains how the tool is used and compares it to several similar tools. ANITA has already been used in logic courses and an evaluation of the tool is presented. 
\end{abstract}

\section{Introduction}

Logic in Computer Science course is part of most Information and Communication Technology curricula, such as the curricula of the Information Systems, Software Engineering, Computer Science, and Computer Engineering of the Federal University of Ceará at Quixadá as mandatory components. The course has a high failure rate. For better assimilation of the contents, it is essential that the students exercise and that they get feedback on the correctness of their proofs.

Many deductive systems are used for teaching the formal reasoning of proofs, such as Axiomatic Systems (a la Hilbert), Natural Deduction System \cite{gentzen1969collected, huth2004logic}, and Analytic Tableaux \cite{smullyan1995first,d1999handbook}.

The Analytic Tableau system is widely used for teaching proofs and appears in many Logic textbooks such as \cite{smullyan1995first, Finger2006, de2008logica, Fitting1996}. This work presents a proof assistant, ANITA (Analytic Tableau Proof Assistant), in order to assist in the teaching-learning of undergraduate and graduate students. For the purpose of teaching deduction systems, we take into account in ANITA the following features: the students should write their proofs as similar as possible to what is available in the textbooks and to what the students would usually write on paper; the tool should be easy-to-use and reduce the number of clicks since mouse-clicking can be tedious; the tool should allow the student to make mistakes and point out errors on the proofs. 

The rest of the paper is organized as follows. We provide a concise definition of Analytic Tableau system in Section~\ref{secao:TA}; We propose to write proofs in Analytic Tableaux in Fitch-style in Section~\ref{secao:ANITA-Fitch}. Section~\ref{secao:ANITA} describes ANITA. Section~\ref{sec:trabalhos-correlatos} compares ANITA to other proof assistants. The evaluation of ANITA is presented in Section~\ref{secao:avaliacao}. And, we conclude this work in Section~\ref{sec:conclusao}.

\section{Analytic Tableaux}
\label{secao:TA}
We now describe Analytic Tableaux for propositional logic which we will subsequently extend to first-order logic.

Analytic Tableaux is an inference method based on \emph{refutation}: to prove $\Gamma\vdash\varphi$, we assert that each formula of $\Gamma$ is \emph{true} and $\varphi$ is \emph{false}, in order to derive a \emph{contradiction}. On the other hand, if no contradiction is obtained, then we construct a \emph{countermodel}, that is, a valuation\footnote{A valuation function $v$ is a mapping from the atoms to the set $\{T,F\}$.} that satisfies $\Gamma$ and does not satisfy $\varphi$.

In the method of analytic tableaux, we define $T~\varphi$ and $F~\varphi$ as signed formulas to stand that $\varphi$ is true and $\varphi$ is false. The first step to constructing a tableau is to label all formulas in $\Gamma$ with $T$ and the formula $\varphi$ with $F$. Starting from the initial tableau, tableau expansion rules can be used to: add new formulas to the end of a branch ($\alpha$-type rules); or split a branch into two branches ($\beta$-type rules). The rules for the construction of tableaux are as follows:
\begin{center}\begin{tabular}{|c|c|c|c|cc|}
  \hline
  $\alpha$ rule
 & \Tree
    [.{ $T~ \varphi\wedge\psi$}
      [.{   $T~\varphi$ \\
            $T~\psi$ }
      ]
    ]
 & \Tree
    [.{ $F~ \varphi\vee\psi$}
      [.{   $F~\varphi$ \\
            $F~\psi$ }
      ]
    ]
 & \Tree
    [.{ $F~ \varphi\rightarrow\psi$}
      [.{   $T~\varphi$ \\
            $F~\psi$ }
      ]
    ]
 & \Tree
    [.{ $T~ \lnot\varphi$}
      [.{   $F~\varphi$ }
      ]
    ]
 & 
   \Tree
    [.{ $F~ \lnot\varphi$}
      [.{   $T~\varphi$}
      ]
    ] 
    
 \\
  \hline
  $\beta$ rule
 & \Tree
    [.{ $F~ \varphi\wedge\psi$}
      [.{   $F~\varphi$}
      ]
      [.{ $F~\psi$ }
      ]
    ]
 & \Tree
    [.{ $T~ \varphi\vee\psi$}
      [.{   $T~\varphi$}
      ]
      [.{ $T~\psi$ }
      ]
    ]
 & \Tree
    [.{ $T~ \varphi\rightarrow\psi$}
      [.{   $F~\varphi$}
      ]
      [.{ $T~\psi$ }
      ]
    ]
 & & 
    \\
  \hline
\end{tabular}
\end{center}

\begin{figure}[ht]
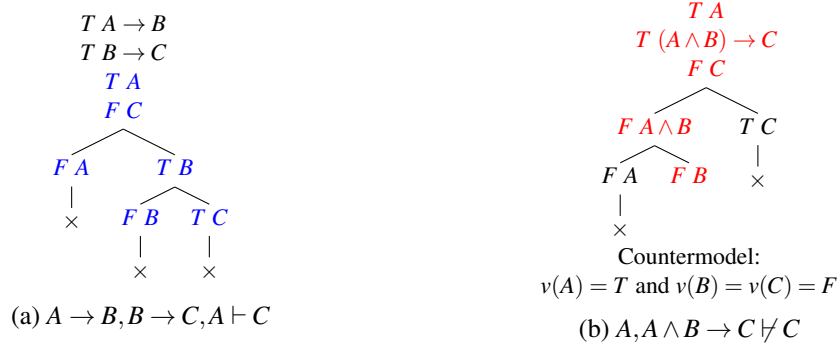

    \centering
\begin{subfigure}{0.45\textwidth}
\footnotesize
        \Tree [.{$T~A\rightarrow B$ \\ $T~B\rightarrow C$ \\ \color{blue}{$T~A$} \\ \color{blue}{$F~C$}} [.{\color{blue}{$F~A$}} [.{$\times$} ] ] [.{\color{blue}{$T~B$}} [.{\color{blue}{$F~B$}} [.{$\times$} ] ] [.{\color{blue}{$T~C$}} [.{$\times$} ] ] ] ]
        \caption{$A\rightarrow B, B\rightarrow C, A\vdash C$}
    \label{fig:TA-TRANSITIVIDADE}
\end{subfigure}
\begin{subfigure}{0.45\textwidth}
\footnotesize
        \Tree [.{\color{red}{$T~A$} \\ \color{red}{$T~(A\land B)\rightarrow C$} \\ \color{red}{$F~C$}} [.{\color{red}{$F~A\land B$}} [.{$F~A$} [.{$\times$} ] ] [.{\color{red}{$F~B$}} ] ] [.{$T~C$} [.{$\times$} ] ] ]        
    \\ \centering Countermodel: \\ $v(A)=T$ and $v(B)=v(C)=F$ %
        \caption{$A, A\wedge B\rightarrow C\not\vdash C$}
    \label{fig:TA-CONTRA-EXEMPLO-1}
\end{subfigure}
\caption{Examples of proofs in Analytic Tableau}
\end{figure}

In each branch, a formula can only be expanded once. A branch that has no more formulas to expand is said to be \textbf{saturated}. A branch that has a pair of formulas $T~\varphi$ and $F~\varphi$ is said to be \textbf{closed}. A closed branch no longer needs to be expanded. A tableau is said to be closed whether it has all its branches closed, i.e., $\Gamma\vdash\varphi$. A saturated and unclosed branch provides a \emph{countermodel}, i.e., $\Gamma\not\vdash\varphi$. Figure~\ref{fig:TA-TRANSITIVIDADE} shows that $A\rightarrow B, B\rightarrow C, A$ entails $C$ as an analytic tableau proof. We use $\times$ as a symbol to close a branch by the signed formulas in the blue nodes. Figure~\ref{fig:TA-CONTRA-EXEMPLO-1} shows a proof, in which we have one of the branches (red nodes) that is saturated. So, we can extract a countermodel from the truth values of the atoms in the branch.

We extend the analytic tableau system in order to include proofs of first-order logic, in which we have all the rules of propositional logic and add the following rules:
\begin{center}\begin{tabular}{|c|c|c|}
  \hline
    $\gamma$ rule
 & \Tree
    [.{ $T~ \forall x\varphi$}
      [.{   $T~\varphi^x_t$ }
      ]
    ]
 & \Tree
    [.{ $F~ \exists x\varphi$}
      [.{   $F~\varphi^x_t$ }
      ]
    ]
 \\
    & \textrm{$t$ is substitutable for $x$ in $\varphi$}
    & \textrm{$t$ is substitutable for $x$ in $\varphi$}
 \\
  \hline
  $\delta$ rule
 & \Tree
    [.{ $F~ \forall x\varphi$}
      [.{   $F~\varphi^x_a$ }
      ]
    ]
 & \Tree
    [.{ $T~ \exists x\varphi$}
      [.{   $T~\varphi^x_a$ }
      ]
    ]
 \\
    & \textrm{a is a new variable}
    & \textrm{a is a new variable}
    \\
  \hline
\end{tabular}
\end{center}

Here $\varphi^x_t$ is the expression obtained from the formula $\varphi$ by replacing the variable $x$, whenever it occurs free in $\varphi$, by the term $t$. For instance, $(H(x)\rightarrow \forall x M(x))^x_y=(H(y)\rightarrow \forall x M(x))$. We can say that a term $t$ is substitutable for $x$ in $\varphi$ if there is no variable $y$ in $t$ that is captured by a $\forall y$ (or $\exists y$) quantifier of $\varphi^x_t$. For example, term $z$ is substitutable for $y$ in $\forall x P(x,y)$. On the other hand, term $x$ is not substitutable for $y$ in $\forall x P(x,y)$.

\begin{figure}[!htb]
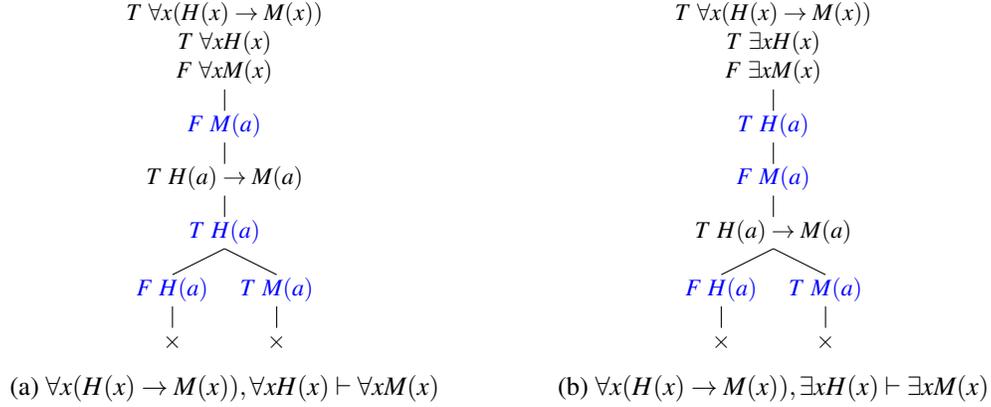

    \centering
\begin{subfigure}{0.45\textwidth}
\footnotesize
        \Tree [.{$T~\forall x (H(x)\rightarrow M(x))$ \\ $T~\forall x H(x)$ \\ $F~\forall x M(x)$} [.{\color{blue}{$F~M(a)$}} [.{$T~H(a)\rightarrow M(a)$} [.{\color{blue}{$T~H(a)$}} [.{\color{blue}{$F~H(a)$}} [.{$\times$} ] ] [.{\color{blue}{$T~M(a)$}} [.{$\times$} ] ] ] ] ] ]
        \caption{$\forall x (H(x)\rightarrow M(x)), \forall x H(x) \vdash \forall x M(x)$}
    \label{fig:TA-FIRST-ORDER1}
\end{subfigure}
\begin{subfigure}{0.45\textwidth}
\footnotesize
        \Tree [.{$T~\forall x (H(x)\rightarrow M(x))$ \\ $T~\exists x H(x)$ \\ $F~\exists x M(x)$} [.{\color{blue}{$T~H(a)$}} [.{\color{blue}{$F~M(a)$}} [.{$T~H(a)\rightarrow M(a)$} [.{\color{blue}{$F~H(a)$}} [.{$\times$} ] ] [.{\color{blue}{$T~M(a)$}} [.{$\times$} ] ] ] ] ] ]
        \caption{$\forall x (H(x)\rightarrow M(x)), \exists x H(x) \vdash \exists x M(x)$}
    \label{fig:TA-FIRST-ORDER2}
\end{subfigure}
\caption{Examples of Proofs in Analytic Tableaux}
\label{fig:TA-FIRST-ORDER}
\end{figure}

The above rules can occur more than once in each branch, as we can make arbitrary substitutions of variables. Thus, in the general case, we will not be able to generate a countermodel. Figure \ref{fig:TA-FIRST-ORDER} shows examples of proofs in Analytic Tableaux.

\section{Analytic Tableaux in Fitch-Style}
\label{secao:ANITA-Fitch}
A (signed) tableau is a certain kind of binary, labeled ordered tree where each node is labeled by a signed formula. However, we can present a version of the analytic tableaux in Fitch-style. The proof is written in a linear and sequential order, in which we number all the lines, and write a statement (signed formula) with its justification which can be a premise, the conclusion of the proof, or the application of one of the inference rules. Each split of a branch is delimited by $\{$ and $\}$. A formula can only be used in a proof at a given point if that formula happened previously and within that branch. In the sequel, we will present all the rules in Fitch-style.

\paragraph{Initial Tableau Rule:}
A proof of $\varphi_1,\varphi_2,\ldots,\varphi_n\vdash\psi$ starts with initial tableau as shown in Figure~\ref{fig:TA-RULE-INITIAL-TABLEAU}, where:
\begin{itemize}
    \item The premises $\varphi_1,\varphi_2,\ldots,\varphi_n$ are represented in one line each, following a sequential numbering, labeled as $T$ (True) and as justification ``Premise''.
    \item The conclusion $\psi$ is defined on the line after the last premise, labeled by $F$ (False) and with the justification ``Conclusion''.
\end{itemize} 

\begin{figure}[!htb]
    \centering
\begin{subfigure}{0.45\textwidth}
    \centering
        \begin{tabular}{lll}
             1. & $~T~\varphi_1$ &~~Premise  \\
             2. &$~T~\varphi_2$ &~~Premise  \\
             $\vdots$ &~~$\vdots$ &~~$\vdots$  \\
             n. &$~T~\varphi_n$ &~~Premise  \\
             n+1. &$~F~\psi$ &~~Conclusion  \\
             $\vdots$ &~~$\vdots$ &~~$\vdots$  \\
        \end{tabular}  
        \caption{Initial Tableau}
    \label{fig:TA-RULE-INITIAL-TABLEAU}
\end{subfigure}
\begin{subfigure}{0.45\textwidth}
        \begin{logicproof}{2}
            T~A & Premise\\
            F~A & Conclusion\\
            \vdots & \vdots
        \end{logicproof}
        \caption{Sample $A\vdash A$}
    \label{fig:TA-RULE-INITIAL-TABLEAU-EX}
\end{subfigure}
\caption{Initial Tableau Rule}
\end{figure}

\paragraph{Closed Branch Rule:} We say that a \textbf{branch is closed} (contains a contradiction $\bot$) in line $p$ if a formula $\varphi$ is labeled in one line $m$ with $T$ and in another line $n$ with $F$ (both before $p$). A closed branch can no longer be expanded. Figure~\ref{fig:TA-RULE-CLOSED} presents the scheme of this rule. Figure~\ref{fig:TA-RULE-CLOSED-EX} shows the proof of $A\vdash A$, in which we close the (single) branch in line $3$ from the formula $A$ referenced in lines $1$ and $2$ as $T$ and $F$, respectively.

\begin{figure}[!htb]
    \centering
\begin{subfigure}{0.45\textwidth}
    \centering
    \begin{tabular}{lll}
         \vdots &~~\vdots &~~\vdots  \\
         $m$. &$~T~\varphi$ &  \\
         \vdots &~~\vdots &~~\vdots  \\
         $n$. &$~F~\varphi$ &  \\
         \vdots &~~\vdots &~~\vdots  \\
         $p$. &~~$\bot$ & ~~$m$,$n$  \\
    \end{tabular}         
        \caption{Closed Branch}
    \label{fig:TA-RULE-CLOSED}
\end{subfigure}
\begin{subfigure}{0.45\textwidth}
        \begin{logicproof}{2}
            T~A & Premise\\
            F~A & Conclusion\\
            \bot & 1,2
        \end{logicproof}
        \caption{$A\vdash A$}
    \label{fig:TA-RULE-CLOSED-EX}
\end{subfigure}
\caption{Closed Branch Rule}
\end{figure}

The \textbf{negation-true rule ($\lnot T$)} is shown in Figure~\ref{fig:TA-NEGATION-TRUE}, where the signed formula $F~\varphi$ can be obtained in line $n$ by the signed formula $T~\lnot\varphi$ in line $m$. In a similar way, the \textbf{negation-false rule ($\lnot F$)}, see Figure~\ref{fig:TA-NEGATION-FALSE}, can be used to show  $T~\varphi$ in line $n$ by $F~\lnot\varphi$ in line $m$. As we can see, in Figure \ref{fig:TA-NEGATION-EX}, we conclude $T~\lnot A$ in line $3$ by the negation-false rule in line $2$. So, we apply $\lnot T$ rule and get $F~A$ in line $4$. Thus, we close the branch in line $5$ (a contradiction), because we have $T~A$ in line $1$ and $F~A$ in line $4$.
\begin{figure}[!htb]
    \centering
\begin{subfigure}{0.25\textwidth}
    \centering
    \begin{tabular}{lll}
         \vdots &~~\vdots &~~\vdots  \\
         m. &$~T~\lnot\varphi$ &  \\
         \vdots &~~\vdots &~~\vdots  \\
         n. &$~F~\varphi$ & ~~$m$  \\
    \end{tabular}    
        \caption{Negation-True ($\lnot T$)}
    \label{fig:TA-NEGATION-TRUE}
\end{subfigure}
\begin{subfigure}{0.25\textwidth}
    \centering
    \begin{tabular}{lll}
         \vdots &~~\vdots &~~\vdots  \\
         m. &$~F~\lnot\varphi$ &  \\
         \vdots &~~\vdots &~~\vdots  \\
         n. &$~T~\varphi$ & ~~$m$  \\
    \end{tabular}            
        \caption{Negation-False ($\lnot F$)}
    \label{fig:TA-NEGATION-FALSE}
\end{subfigure}
\begin{subfigure}{0.40\textwidth}
        \begin{logicproof}{2}
            T~A & Premise\\
            F~\lnot\lnot A & Conclusion\\
            T~\lnot A & 2\\
            F~A & 3\\
            \bot & 1,4
        \end{logicproof}
        \caption{$A\vdash \lnot\lnot A$}
    \label{fig:TA-NEGATION-EX}
\end{subfigure}
\caption{Negation Rules}
\end{figure}

The \textbf{and-true rule ($\land T$)} is shown in Figure~\ref{fig:TA-RULE-CONJUNCTION-TRUE}, in which the signed formulas $T~\varphi$ and $T~\psi$ in lines $n$ and $n+1$, respectively, are obtained by the signed formula $T~\varphi\land\psi$ in line $m$. For example, in Figure~\ref{fig:TA-RULE-CONJUNCTION-TRUE-EX}, we apply rule $\land T$ to $T~A\land B$ in line $1$ and derive $T~A$ and $T~B$, in lines $3$ and $4$.

\begin{figure}[!htb]
    \centering
\begin{subfigure}{0.45\textwidth}
    \centering
    \begin{tabular}{lll}
         \vdots &~~\vdots &~~\vdots  \\
         m. &$~T~\varphi\land\psi$ &  \\
         \vdots &~~\vdots &~~\vdots  \\
         n. &$~T~\varphi$ & ~~$m$  \\
         n+1. &$~T~\psi$ & ~~$m$  \\
    \end{tabular}    
        \caption{And-True ($\land T$)}
    \label{fig:TA-RULE-CONJUNCTION-TRUE}
\end{subfigure}
\begin{subfigure}{0.45\textwidth}
        \begin{logicproof}{2}
            T~A\land B & Premise\\
            F~A & Conclusion\\
            T~A & 1\\
            T~B & 1\\
            \bot & 2,3
        \end{logicproof}
        \caption{$A\land B\vdash A$}
    \label{fig:TA-RULE-CONJUNCTION-TRUE-EX}
\end{subfigure}
\caption{And-True Rule}
\end{figure}

The \textbf{and-false rule ($\land F$)} is shown in Figure~\ref{fig:TA-RULE-CONJUNCTION-FALSE}. This rule is applied to $F~\varphi\land\psi$ in line $m$ and splits this branch into two: one that starts in line $n$ with $F~\varphi$; and the other in line $p$ with $F~\psi$. To delimit the respective branches, we use the symbols $\{$ and $\}$. For instance, in Figure~\ref{fig:TA-RULE-CONJUNCTION-FALSE-EX}, we apply $\land F$ rule to $F~A\wedge B$ in line $3$ and split this branch: 
\begin{enumerate}
    \item In the branch starting in line $4$ with $F~A$ which is used with $T~A$ from line $1$ to close this branch in line $5$.
    \item In the branch starting in line $6$ with $F~B$ which is used with $T~B$ from line $2$ to close this branch in line $7$.
\end{enumerate}

It is worth noting that we can only reference the formulas in the same branch. So, for example, the formula $F~A$ in line $4$ could not be referenced in the branch starting in line $6$.

\begin{figure}[!htb]
    \centering
\begin{subfigure}{0.45\textwidth}
    \centering
    \begin{tabular}{lll}
         \vdots &~~\vdots &~~\vdots  \\
         m. &$~F~\varphi\land \psi$ &  \\
         \vdots &~~\vdots &~~\vdots  \\
         n. &\{\qquad $F~\varphi$ & ~~$m$  \\
         \vdots &\qquad~\vdots &~~\vdots  \\
          &\} &  \\
         p. &\{\qquad $F~\psi$ & ~~$m$  \\
         \vdots &\qquad~\vdots &~~\vdots  \\
          &\} &  \\
    \end{tabular}    
        \caption{And-False ($\land F$)}
    \label{fig:TA-RULE-CONJUNCTION-FALSE}
\end{subfigure}
\begin{subfigure}{0.45\textwidth}
        \begin{logicproof}{2}
            T~A & Premise\\
            T~B & Premise\\
            F~A\land B & Conclusion\\
            \{\qquad F~A & 3\\
            ~\qquad \bot & 1,4$~~\}$\\
            \{\qquad F~B & 3\\
            ~\qquad \bot & 2,6$~~\}$
        \end{logicproof}
        \caption{$A,B\vdash A\land B$}
    \label{fig:TA-RULE-CONJUNCTION-FALSE-EX}
\end{subfigure}
\caption{And-False Rule}
\end{figure}

The \textbf{or-true rule ($\lor T$)} is presented in Figure~\ref{fig:TA-RULE-DISJUNCTION-TRUE}. We apply this rule to $T~\varphi\lor\psi$ in line $m$ and we split this branch into two new branches: one that starts in line $n$ with $T~\varphi$; and the other in line $p$ with $T~\psi$. For example, in Figure~\ref{fig:TA-RULE-DISJUNCTION-TRUE-EX}, the rule $\lor T$ is applied to $T~A\lor B$ in line $1$ and we split this branch:
\begin{enumerate}
    \item In the branch that starts in line $5$ with $T~A$ which is used with $F~A$ in line $3$ to close this branch in line $6$.
    \item In the branch starting in line $7$ with $T~B$ which is used with $F~B$ in line $4$ to close this branch in line $8$.
\end{enumerate}

\begin{figure}[!htb]
    \centering
\begin{subfigure}{0.45\textwidth}
    \centering
    \begin{tabular}{lll}
         \vdots &~~\vdots &~~\vdots  \\
         m. &$~T~\varphi\lor\psi$ &  \\
         \vdots &~~\vdots &~~\vdots  \\
         n. &\{\qquad $T~\varphi$ & ~~$m$  \\
         \vdots &\qquad~\vdots &~~\vdots  \\
          &\} &  \\
         p. &\{\qquad $T~\psi$ & ~~$m$  \\
         \vdots &\qquad~\vdots &~~\vdots  \\
          &\} &  \\
    \end{tabular}    
        \caption{Or-True ($\lor T$)}
    \label{fig:TA-RULE-DISJUNCTION-TRUE}
\end{subfigure}
\begin{subfigure}{0.45\textwidth}
        \begin{logicproof}{2}
            T~A\lor B & Premise\\
            T~\lnot B & Premise\\
            F~A & Conclusion\\
            F~B & 2\\
            \{\qquad T~A & 1\\
            ~\qquad \bot & 3,5$~~\}$\\
            \{\qquad T~B & 1\\
            ~\qquad \bot & 4,7$~~\}$
        \end{logicproof}
        \caption{$A\lor B, \lnot B\vdash A$}
    \label{fig:TA-RULE-DISJUNCTION-TRUE-EX}
\end{subfigure}
\caption{Or-True Rule}
\end{figure}

The \textbf{or-false rule ($\lor F$)} is shown in Figure~\ref{fig:TA-RULE-DISJUNCTION-FALSE}, in which the signed formulas $F~\varphi$ and $F~\psi$ in lines $n$ and $n+1$, respectively, are obtained by the signed formula $F~\varphi\lor\psi$. For example, in Figure~\ref{fig:TA-RULE-DISJUNCTION-FALSE-EX}, we apply rule $\lor F$ to $F~A\lor B$ in line $2$ and derive $F~A$ and $F~B$, in lines $3$ and $4$.

\begin{figure}[!htb]
    \centering
\begin{subfigure}{0.45\textwidth}
    \centering
    \begin{tabular}{lll}
         \vdots &~~\vdots &~~\vdots  \\
         m. &$~F~\varphi\lor\psi$ &  \\
         \vdots &~~\vdots &~~\vdots  \\
         n. &$~F~\varphi$ & ~~$m$  \\
         n+1. &$~F~\psi$ & ~~$m$  \\
    \end{tabular}    
        \caption{Or-False ($\lor F$)}
    \label{fig:TA-RULE-DISJUNCTION-FALSE}
\end{subfigure}
\begin{subfigure}{0.45\textwidth}
        \begin{logicproof}{2}
            T~A & Premise\\
            F~A\lor B & Conclusion\\
            F~A & 2\\
            F~B & 2\\
            \bot & 1,3
        \end{logicproof}
        \caption{$A\vdash A\lor B$}
    \label{fig:TA-RULE-DISJUNCTION-FALSE-EX}
\end{subfigure}
\caption{Or-False Rule}
\end{figure}

The \textbf{implication-true rule ($\rightarrow T$)} is presented in Figure~\ref{fig:TA-RULE-IMPLICATION-TRUE}. We apply this rule to $T~\varphi\rightarrow\psi$ in line $m$ and we split this branch into two new branches: one that starts in line $n$ with $F~\varphi$; and the other in line $p$ with $T~\psi$. For example, in Figure~\ref{fig:TA-RULE-IMPLICATION-TRUE-EX}, the rule $\rightarrow T$ is applied to $T~A\rightarrow B$ in line $1$ and we split this branch:
\begin{enumerate}
    \item In the branch that starts in line $5$ with $F~\lnot A$, in which we apply $\lnot F$ to obtain $T~A$ in line $6$ and, then, we use $F~A$ in line $3$ to close this branch in line $7$.
    \item In the branch starting in line $8$ with $T~B$ which is used with $F~B$ in line $4$ to close this branch in line $9$.
\end{enumerate}

\begin{figure}[!htb]
    \centering
\begin{subfigure}{0.45\textwidth}
    \centering
    \begin{tabular}{lll}
         \vdots &~~\vdots &~~\vdots  \\
         m. &$~T~\varphi\rightarrow\psi$ &  \\
         \vdots &~~\vdots &~~\vdots  \\
         n. &\{\qquad $F~\varphi$ & ~~$m$  \\
         \vdots &\qquad~\vdots &~~\vdots  \\
          &\} &  \\
         p. &\{\qquad $T~\psi$ & ~~$m$  \\
         \vdots &\qquad~\vdots &~~\vdots  \\
          &\} &  \\
    \end{tabular}    
        \caption{Implication-True ($\rightarrow T$)}
    \label{fig:TA-RULE-IMPLICATION-TRUE}
\end{subfigure}
\begin{subfigure}{0.45\textwidth}
        \begin{logicproof}{2}
            T~\lnot A\rightarrow B & Premise\\
            F~A\lor B & Conclusion\\
            F~A & 2\\
            F~B & 2\\
            \{\qquad F~\lnot A & 1\\
            ~\qquad T~A & 5\\
            ~\qquad \bot & 6,3$~~\}$\\
            \{\qquad T~B & 1\\
            ~\qquad \bot & 8,4$~~\}$
        \end{logicproof}
        \caption{$\lnot A\rightarrow B\vdash A\lor B$}
    \label{fig:TA-RULE-IMPLICATION-TRUE-EX}
\end{subfigure}
\caption{Implication-True Rule}
\end{figure}

The \textbf{implication-false rule ($\rightarrow F$)} is shown in Figure~\ref{fig:TA-RULE-IMPLICATION-FALSE}, in which the signed formulas $T~\varphi$ and $F~\psi$ in lines $n$ and $n+1$, respectively, are obtained by the signed formula $F~\varphi\rightarrow\psi$. For example, in Figure~\ref{fig:TA-RULE-IMPLICATION-FALSE-EX}, we apply rule $\rightarrow F$ to $F~A\rightarrow B$ in line $2$ and derive $T~A$ and $F~B$, in lines $3$ and $4$.

\begin{figure}[!htb]
    \centering
\begin{subfigure}{0.45\textwidth}
    \centering
    \begin{tabular}{lll}
         \vdots &~~\vdots &~~\vdots  \\
         m. &$~F~\varphi\rightarrow\psi$ &  \\
         \vdots &~~\vdots &~~\vdots  \\
         n. &$~T~\varphi$ & ~~$m$  \\
         n+1. &$~F~\psi$ & ~~$m$  \\
    \end{tabular}    
    \caption{Implication-False ($\rightarrow F$)}
    \label{fig:TA-RULE-IMPLICATION-FALSE}
\end{subfigure}
\begin{subfigure}{0.45\textwidth}
        \begin{logicproof}{2}
            T~B & Premise\\
            F~A\rightarrow B & Conclusion\\
            T~A & 2 \\
            F~B & 2\\
            \bot & 1,4
        \end{logicproof}
        \caption{$B\vdash A\rightarrow B$}
    \label{fig:TA-RULE-IMPLICATION-FALSE-EX}
\end{subfigure}
\caption{Implication-False Rule}
\end{figure}

The \textbf{universal-true rule ($\forall T$)} is shown in Figure~\ref{fig:TA-RULE-FORALL-TRUE}, in which we apply $\forall T$-rule to signed formula $T~\forall x\varphi(x)$ in line $m$ and obtain, in line $n$, a signed formula $T~\varphi^x_t$, where $t$ is substitutable for $x$ in $\varphi$. Figure~\ref{fig:TA-RULE-FORALL-TRUE-EX} shows the use of this rule to $T~\forall x(H(x)\rightarrow M(x))$ in line $1$ to get $T~H(s)\rightarrow M(s)$ in line $4$.

\begin{figure}[!htb]
    \centering
\begin{subfigure}{0.45\textwidth}
    \centering
    \begin{tabular}{lll}
         \vdots &~~\vdots &~~\vdots  \\
         m. &$~T~\forall x\varphi$ &  \\
         \vdots &~~\vdots &~~\vdots  \\
         n. &$~T~\varphi^x_t$ & ~~$m$  \\
    \end{tabular}    
    \\ \textrm{$t$ is substitutable for $x$ in $\varphi$}
    \caption{Universal-True ($\forall T$)}
    \label{fig:TA-RULE-FORALL-TRUE}
\end{subfigure}
\begin{subfigure}{0.45\textwidth}
        \begin{logicproof}{2}
            T~\forall x(H(x) \rightarrow M(x)) & Premise\\
            T~H(s) & Premise\\
            F~M(s) & Conclusion\\
            T~H(s) \rightarrow M(s) & 1\\
            \{\qquad F~H(s) & 4\\
            ~\qquad \bot & 2,5$~~\}$\\
            \{\qquad T~M(s) & 4\\
            ~\qquad \bot & 7,3$~~\}$
        \end{logicproof}
        \caption{$\forall x(H(x) \rightarrow M(x)), H(s)\vdash M(s)$}
    \label{fig:TA-RULE-FORALL-TRUE-EX}
\end{subfigure}
\caption{Universal-True Rule}
\end{figure}

The \textbf{universal-false rule ($\forall F$)} is shown in Figure~\ref{fig:TA-RULE-FORALL-FALSE}, in which we apply $\forall F$-rule to signed formula $F~\forall x\varphi(x)$ in line $m$ and obtain, in line $n$, a signed formula $F~\varphi^x_a$, where $a$ is a new variable. Figure~\ref{fig:TA-RULE-FORALL-FALSE-EX} illustrates the use of this rule to $F~\forall x M(x)$ in line $3$ to conclude $F~M(a)$ in line $4$.

\begin{figure}[!htb]
    \centering
\begin{subfigure}{0.45\textwidth}
    \centering
    \begin{tabular}{lll}
         \vdots &~~\vdots &~~\vdots  \\
         m. &$~F~\forall x\varphi$ &  \\
         \vdots &~~\vdots &~~\vdots  \\
         n. &$~F~\varphi^x_a$ & ~~$m$  \\
    \end{tabular}    
    \\ \textrm{$a$ is new variable}
    \caption{Universal-False ($\forall F$)}
    \label{fig:TA-RULE-FORALL-FALSE}
\end{subfigure}
\begin{subfigure}{0.45\textwidth}
        \begin{logicproof}{2}
            T~\forall x(H(x) \rightarrow M(x)) & Premise\\
            T~\forall x H(x) & Premise\\
            F~\forall x M(x) & Conclusion\\
            F~M(a) & 3\\
            T~H(a) & 2\\
            T~H(a) \rightarrow M(a) & 1\\
            \{\qquad F~H(a) & 6\\
            ~\qquad \bot & 5,7$~~\}$\\
            \{\qquad T~M(a) & 6\\
            ~\qquad \bot & 9,4$~~\}$
        \end{logicproof}
        \caption{$\forall x(H(x) \rightarrow M(x)), \forall x H(x)\vdash \forall x M(x)$}
    \label{fig:TA-RULE-FORALL-FALSE-EX}
\end{subfigure}
\caption{Universal-False Rule}
\end{figure}

The \textbf{existential-true rule ($\exists T$)} is shown in Figure~\ref{fig:TA-RULE-EXISTS-TRUE}, in which we apply $\exists T$-rule to signed formula $T~\exists x\varphi(x)$ in line $m$ and obtain, in line $n$, a signed formula $T~\varphi^x_a$, where $a$ is a new variable. Figure~\ref{fig:TA-RULE-EXISTS-TRUE-EX} illustrates the use of this rule to $T~\exists x~H(x)$ in line $2$ to get $T~H(a)$ in line $4$.

\begin{figure}[!htb]
    \centering
\begin{subfigure}{0.45\textwidth}
    \centering
    \begin{tabular}{lll}
         \vdots &~~\vdots &~~\vdots  \\
         m. &$~T~\exists x\varphi$ &  \\
         \vdots &~~\vdots &~~\vdots  \\
         n. &$~T~\varphi^x_a$ & ~~$m$  \\
    \end{tabular}    
    \\ \textrm{$a$ is new variable}
    \caption{Existential-True ($\exists T$)}
    \label{fig:TA-RULE-EXISTS-TRUE}
\end{subfigure}
\begin{subfigure}{0.45\textwidth}
        \begin{logicproof}{2}
            T~\forall x(H(x) \rightarrow M(x)) & Premise\\
            T~\exists x H(x) & Premise\\
            F~ \exists x M(x) & Conclusion\\
            T~H(a) & 2\\
            F~M(a) & 3\\
            T~H(a) \rightarrow M(a)  & 1\\
            \{\qquad F~H(a) & 6\\
            ~\qquad \bot & 4,7$~~\}$\\
            \{\qquad T~M(a) & 6\\
            ~\qquad \bot & 5,9$~~\}$
        \end{logicproof}
        \caption{$\forall x(H(x) \rightarrow M(x)), \exists x H(x)\vdash \exists x M(x)$}
    \label{fig:TA-RULE-EXISTS-TRUE-EX}
\end{subfigure}
\caption{Existential-True Rule}
\end{figure}

The \textbf{existential-false rule ($\exists F$)} is shown in Figure~\ref{fig:TA-RULE-EXISTS-FALSE}, in which we apply $\exists F$-rule to signed formula $F~\exists x\varphi(x)$ in line $m$ and obtain, in line $n$, a signed formula $F~\varphi^x_t$, where $t$ is substitutable for $x$ in $\varphi$. Figure~\ref{fig:TA-RULE-EXISTS-FALSE-EX} illustrates the use of this rule to $F~\exists x P(x)$ in line $4$ to conclude $P(a)$ in line $5$.

\begin{figure}[!htb]
    \centering
\begin{subfigure}{0.45\textwidth}
    \centering
    \begin{tabular}{lll}
         \vdots &~~\vdots &~~\vdots  \\
         m. &$~F~\exists x\varphi$ &  \\
         \vdots &~~\vdots &~~\vdots  \\
         n. &$~F~\varphi^x_t$ & ~~$m$  \\
    \end{tabular}    
    \\ \textrm{$t$ is substitutable for $x$ in $\varphi$}
    \caption{Existential-False ($\exists F$)}
    \label{fig:TA-RULE-EXISTS-FALSE}
\end{subfigure}
\begin{subfigure}{0.45\textwidth}
        \begin{logicproof}{2}
            T~P(a) & Premise\\
            T~\exists x P(x) \rightarrow B  & Premise\\
            F~ B & Conclusion\\
            \{\qquad F~\exists x P(x) & 2\\
            ~\qquad  F~P(a) & 4\\
            ~\qquad \bot & 1,5$~~\}$\\
            \{\qquad T~B & 2\\
            ~\qquad \bot & 7,3$~~\}$
        \end{logicproof}
        \caption{$P(a), \exists x P(x) \rightarrow B \vdash B$}
    \label{fig:TA-RULE-EXISTS-FALSE-EX}
\end{subfigure}
\caption{Existential-False Rule}
\end{figure}

\section{Analytic Tableau Proof Assistant (ANITA)}
\label{secao:ANITA}
The ANITA\footnote{ANITA source-code is available at \url{https://github.com/daviromero/anita} under a MIT License.} proof assistant, \textit{Analytic Tableau Proof Assistant}, is a tool written in Python that can be used as a \textit{desktop} application, or in a web platform\footnote{ANITA is available at: \url{https://sistemas.quixada.ufc.br/anita/en/}}. The main idea is that the students can write their proofs as similar as possible to what is available in the textbooks and to what the students would usually write on paper. ANITA allows the students to automatically check whether a proof in the analytic tableaux is valid. If the proof is not correct, the tool will display the errors on the proof. So, the students may make mistakes and learn from the errors. The web interface is very easy-to-use and has: 
\begin{itemize}
    \item An area for editing the proof in plain text. The students should write a proof in Fitch-style presented in Section \ref{secao:ANITA-Fitch}.
    \item A message area to display whether the proof is valid, the countermodel, or the errors on the proof.
    \item And the following \textit{links}: \emph{Check}, to check the correctness of the proof; \emph{Manual}, to view a document with the inference rules and examples; \emph{LaTeX}, to generate the LaTeX code\footnote{Use the \emph{qtree} package in your LaTeX code.} of the trees from a valid proof; \emph{Latex in Overleaf} to open the proof source code directly in Overleaf\footnote{Overleaf is a collaborative platform for editing LaTeX. Available at: \url{http://overleaf.com/}}.
\end{itemize}

To facilitate the writing of the proofs, we made the following conventions in ANITA:
\begin{itemize}
    \item The Atoms\footnote{An atomic formula or atom is simply a predicate applied to a tuple of terms; that is, an atomic formula is a formula of the form $P(t_1,\ldots, t_n)$ for $P$ a predicate, and the $t_n$ terms.} are written in capital letters (e.g. \textrm{A, B,  H(x)});
    \item Variables are written with the first letter in lowercase, followed by letters and numbers (e.g. \textrm{x, x0});
    \item Formulas with $\forall x$ and $\exists x$ are represented by $Ax$ and $Ex$ (`A' and `E' followed by the variable x). For instance, \textrm{Ax (H(x)$-\!\!>$M(x))} represents $\forall x~(H(x)\rightarrow M(x))$.
    \item Figure \ref{fig:equivalencia-ANITA} shows the equivalence of logic symbols and those used in ANITA.
    \item The order of precedence of quantifiers and logical connectives is defined by $\lnot,\forall,\exists,\wedge,\vee,\rightarrow$ with right alignment. For example:
    \begin{itemize}
        \item Formula \textrm{$\sim$A\&B$-\!\!>$C} represents formula $(((\lnot A)\land B)\rightarrow C)$;
        \item The theorem \textrm{$\sim$A$\mid$B $|-$ A$-\!\!>$C} represents $((\lnot A)\vee B)\vdash (A\rightarrow C)$.
    \end{itemize} 
    \item Each inference rule will be named by its respective connective and the truth value of the signed formula. For example, \textrm{\&T} represents the and-true rule. Optionally, the rule name can be omitted.
    \item The justifications for the premises and the conclusion use the reserved words \textrm{pre} and \textrm{conclusion}, respectively.
\end{itemize}
\begin{figure}[H]
\footnotesize
\centering
\begin{tabular}{|c|c|c|c|c|c|c|c|c|c|}
    \hline 
    \textbf{Symbol} &  $\lnot$& $\land$ & $\lor$ & $\rightarrow$& $\forall x$ & $\exists x$ & $\bot$& branch &$\vdash$\\
    \hline 
     \textbf{LaTeX} &  $\backslash\textrm{lnot}$ & $\backslash\textrm{land}$& $\backslash\textrm{lor}$& $\backslash\textrm{rightarrow}$ & $\backslash\textrm{forall x}$ & $\backslash\textrm{exists x}$ & $\backslash\textrm{bot}$  & $[.~]$  & $\backslash\textrm{vdash}$
     \\
    \hline
     \textbf{ANITA} &  $\sim$  & \& & $\mid$ & $-\!\!>$ & Ax & Ex & $\textrm{@}$  & $\{\}$ &$|-$
     \\
    \hline 
\end{tabular}
\caption{Equivalence between the symbols of logic, ANITA and LaTeX}
  \label{fig:equivalencia-ANITA}
\end{figure}

\begin{figure}[H]
    \centering
\begin{subfigure}{0.6\textwidth}
\centering
    \includegraphics[width=\textwidth]{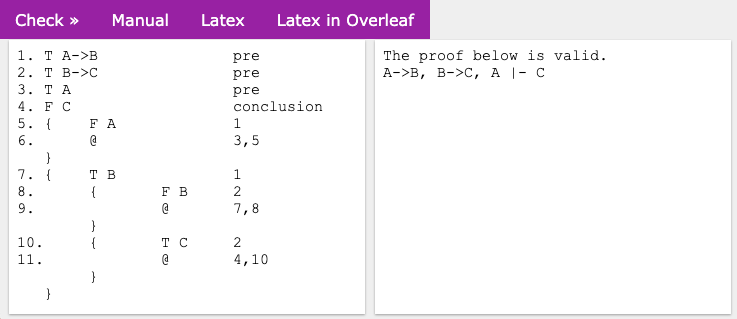}
    \caption{ANITA: $A\rightarrow B, B\rightarrow C, A\vdash C$}
    \label{fig:TA-Transitivivade-anita}
\end{subfigure}
\begin{subfigure}{0.35\textwidth}
\footnotesize\centering
\Tree [.{$T~A\rightarrow B$ \\ $T~B\rightarrow C$ \\ \color{blue}{$T~A$} \\ \color{blue}{$F~C$}} [.{\color{blue}{$F~A$}} [.{$\times$} ] ] [.{\color{blue}{$T~B$}} [.{\color{blue}{$F~B$}} [.{$\times$} ] ] [.{\color{blue}{$T~C$}} [.{$\times$} ] ] ] ]
    \caption{Analytic Tableau Proof}
    \label{fig:TA-Transitivivade-anita-tree}
\end{subfigure}
\begin{subfigure}{0.6\textwidth}
\centering
    \includegraphics[width=\textwidth]{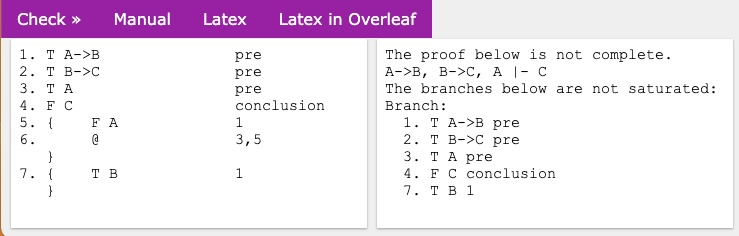}
    \caption{Open Branch in ANITA: $A\rightarrow B, B\rightarrow C, A\vdash C$}
    \label{fig:TA-OPEN-BRANCH-anita}
\end{subfigure}
\begin{subfigure}{0.35\textwidth}
\footnotesize\centering
    \Tree [.{\color{red}{$T~A\rightarrow B$} \\ \color{red}{$T~B\rightarrow C$} \\ \color{red}{$T~A$} \\ \color{red}{$F~C$}} [.{$F~A$} [.{$\times$} ] ] [.{\color{red}{$T~B$}} ] ]
        \caption{Analytic Tableau Proof}
    \label{fig:TA-OPEN-BRANCH-anita-tree}
\end{subfigure}
\caption{ANITA: $A\rightarrow B, B\rightarrow C, A\vdash C$}
\end{figure}

\begin{figure}[H]
    \centering
    \includegraphics[scale=0.35]{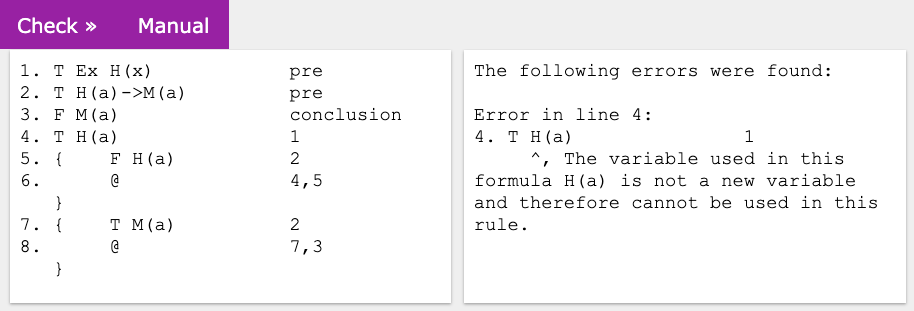}
\caption{ANITA: An error in a proof}
    \label{fig:TA-ERROR-FIRST-ORDER-anita}
\end{figure}

Figure~\ref{fig:TA-Transitivivade-anita} shows a valid proof of $A\rightarrow B, B\rightarrow C, A\vdash C$ in ANITA, and Figure~\ref{fig:TA-Transitivivade-anita-tree} shows the tree generated by ANITA, where the blue nodes (signed formulas) point out the closed branches. Figure~\ref{fig:TA-OPEN-BRANCH-anita} illustrates an example of an incomplete proof of $A\rightarrow B, B\rightarrow C, A\vdash C$ in ANITA, whereas Figure~\ref{fig:TA-OPEN-BRANCH-anita-tree} displays the open branch in red of the analytic tableau that was generated by ANITA.
Figure~\ref{fig:TA-ERROR-FIRST-ORDER-anita} shows a message that the existential-true rule is not applied correctly to signed formula $T~Ex~H(x)$ in line $1$ to obtain $T~H(a)$ in line 4, because the term $a$ is not a new variable (see line $3$).

\begin{figure}[H]
    \centering
\begin{subfigure}{0.55\textwidth}
\centering
    \includegraphics[scale=0.3]{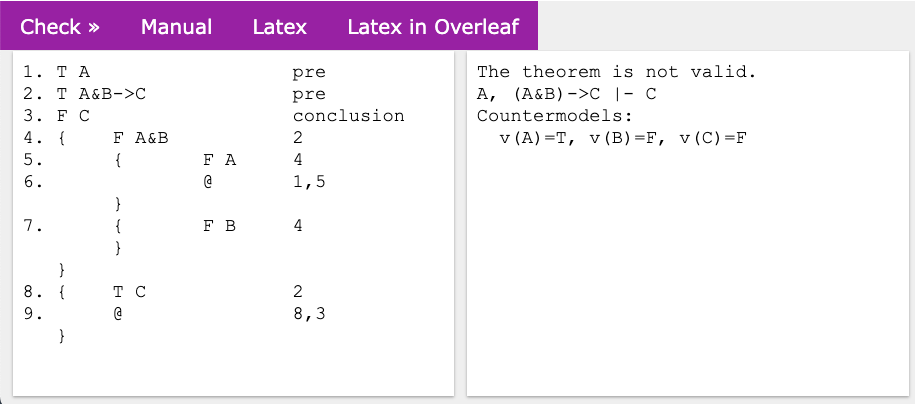}
    \caption{ANITA: $A, A\land B\rightarrow C\not\vdash C$}
    \label{fig:TA-COUNTERMODEL-anita}
\end{subfigure}
\begin{subfigure}{0.4\textwidth}
\footnotesize\centering
        \Tree [.{\color{red}{$T~A$} \\ \color{red}{$T~(A\land B)\rightarrow C$} \\ \color{red}{$F~C$}} [.{\color{red}{$F~A\land B$}} [.{$F~A$} [.{$\times$} ] ] [.{\color{red}{$F~B$}} ] ] [.{$T~C$} [.{$\times$} ] ] ]
        \caption{Analytic Tableau Proof}
    \label{fig:TA-COUNTERMODEL-anita-tree}
\end{subfigure}
\begin{subfigure}{0.55\textwidth}
\centering
    \includegraphics[scale=0.47]{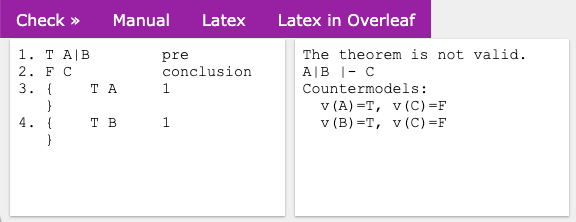}
    \caption{ANITA: $A\lor B\not\vdash C$}
    \label{fig:TA-COUNTERMODEL-anita-2}
\end{subfigure}
\begin{subfigure}{0.4\textwidth}
\footnotesize\centering
        \Tree [.{\color{red}{$T~A\lor B$} \\ \color{red}{$F~C$}} [.{\color{red}{$T~A$}} ] [.{\color{red}{$T~B$}} ] ]
    \caption{Analytic Tableau Proof}
    \label{fig:TA-COUNTERMODEL-anita-tree-2}
\end{subfigure}
\caption{ANITA: Sample of countermodel}
\end{figure}

Figure~\ref{fig:TA-COUNTERMODEL-anita} presents a proof that $A, A\land B\rightarrow C$ does not entail $C$, and ANITA displays the countermodel of the proof. Figure~\ref{fig:TA-COUNTERMODEL-anita-tree} displays the saturated branch in red of the analytic tableau that provides a countermodel that was generated by ANITA. Figure~\ref{fig:TA-COUNTERMODEL-anita-tree-2} displays two saturated branches in red of the analytic tableau that provide countermodels of the proof $A\lor B\not\vdash C$ that were generated by ANITA, see Figure \ref{fig:TA-COUNTERMODEL-anita-2}. Note that in the open branch (lines $1, 2$ and $3$) the atomic formula $B$ does not occur, then $v(B)$ can be $T$ or false $F$, and the countermodel is displayed by $v(A)=T, v(C)=F$.

\section{Related Work}\label{sec:trabalhos-correlatos}

In this article, we focus on Natural Deduction and Analytic Tableau proof assistants. Although, there are proof assistants for other systems, such as SeCaV \cite{EPTCS357.4}. We summarize the features of proof assistants, as well as highlight the similarities and differences between these tools and the proposal in this work. We also provide the proof of $A\rightarrow~B,B\rightarrow~C,A\vdash~C$ in each proof assistant.

\begin{itemize}
    \item The Jape\footnote{Jape source-code is available at \url{https://github.com/RBornat/jape/}} \cite{bornat1996jape} is a desktop proof assistant to write proofs in Fitch-style in Natural Deduction. The proofs are performed by inserting the inference rules through its GUI. 
    \item The ProofWeb \cite{ProofWeb2010} is a web interface that intends to be an evolution of JAPE and uses Coq\footnote{Coq is available at \url{https://coq.inria.fr/}} that is state-of-art proof assistant for writing mathematical proofs. The user must write the proofs in a text area or use the GUI to add the inference rules. The ProofWeb can display proofs in Fitch or Gentzen-styles.
    \item The Panda \cite{Panda2011} is also a desktop proof assistant which differs from the previous ones by allowing the writing of proofs in Gentzen-style from its GUI. 
    \item The NaDeA\footnote{Available at \url{https://nadea.compute.dtu.dk/}} \cite{EPTCS267.9} is a web proof assistant for Natural Deduction with a formalization in Isabelle. The user must write the proofs through its user interface which is based on clicking.
    \item The NADIA\footnote{Available at \url{https://sistemas.quixada.ufc.br/nadia/}}\footnote{NADIA source-code is available at \url{https://github.com/daviromero/nadia} under a MIT License.} \cite{NADIA:22} is a web proof assistant for Natural Deduction, in Fitch-style. NADIA allows students to write their proofs as closely as possible to the proofs they take on paper, by using an input syntax code similar to \cite{huth2004logic}. NADIA displays proofs in Fitch or Gentzen-style. 
    \item The Carnap.io\footnote{Available at \url{https://carnap.io}}  \cite{EPTCS267.5} is a free and open-source Haskell framework for creating and exploring formal languages, logics, and semantics. A web proof assistant for Analytic Tableaux\footnote{Available at \url{https://carnap.io/srv/doc/truth-tree.md}} is available and can be used to construct proofs by using the GUI interface.
    \item The Tree Proof Generator Tableau\footnote{Available  at \url{https://www.umsu.de/trees}} is a tableau prover for classical propositional and first-order logic, as well as some modal logics. The prover is written in Javascript and runs entirely in the browser. The user can enter a formula of standard propositional, predicate, or modal logic and the prover will automatically try to find either a countermodel or a tree proof.  
\end{itemize}

\begin{figure}[H]
\centering
\begin{subfigure}{0.45\textwidth}
    \centering
    \includegraphics[scale=0.35]{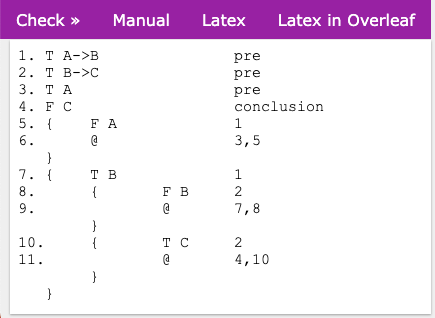}
    \caption{ANITA: $A\rightarrow B, B\rightarrow C, A\vdash C$}
    \label{fig:anita}
\end{subfigure}
\begin{subfigure}{0.4\textwidth}
    \centering
    \includegraphics[scale=.2]{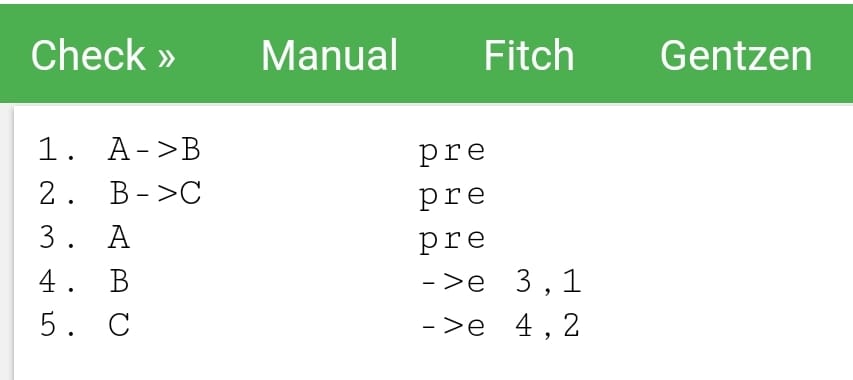}
    \caption{NADIA: $A\rightarrow B, B\rightarrow C, A\vdash C$}
    \label{fig:nadia}
\end{subfigure}
\caption{ANITA and NADIA Proof Assistants}
\end{figure}

ANITA is very similar to NADIA. Both systems receive as input a text with a proof of a theorem and check whether the proof is correct or not. If not, the tools display the errors found. The main difference between the proof assistants is that ANITA accepts proofs in Analytic Tableaux (see Figure~\ref{fig:anita}) and NADIA in Natural Deduction (see Figure~\ref{fig:nadia}). The parser of the proofs in ANITA is completely different from the NADIA parser, as each implements a very different set of inference rules.

\begin{figure}[H]
    \centering
\begin{subfigure}{0.55\textwidth}
    \centering
    \includegraphics[scale=0.4]{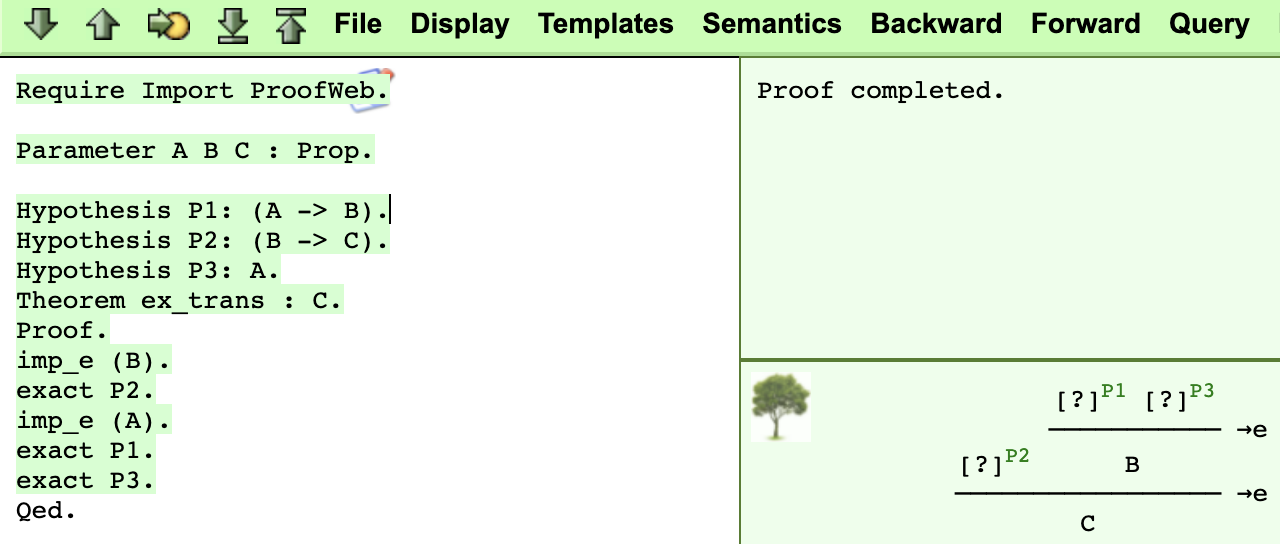}
    \caption{ProofWeb: $A\rightarrow B, B\rightarrow C, A\vdash C$}
    \label{fig:proofweb}
\end{subfigure}
\begin{subfigure}{0.4\textwidth}
    \centering
    \includegraphics[scale=0.35]{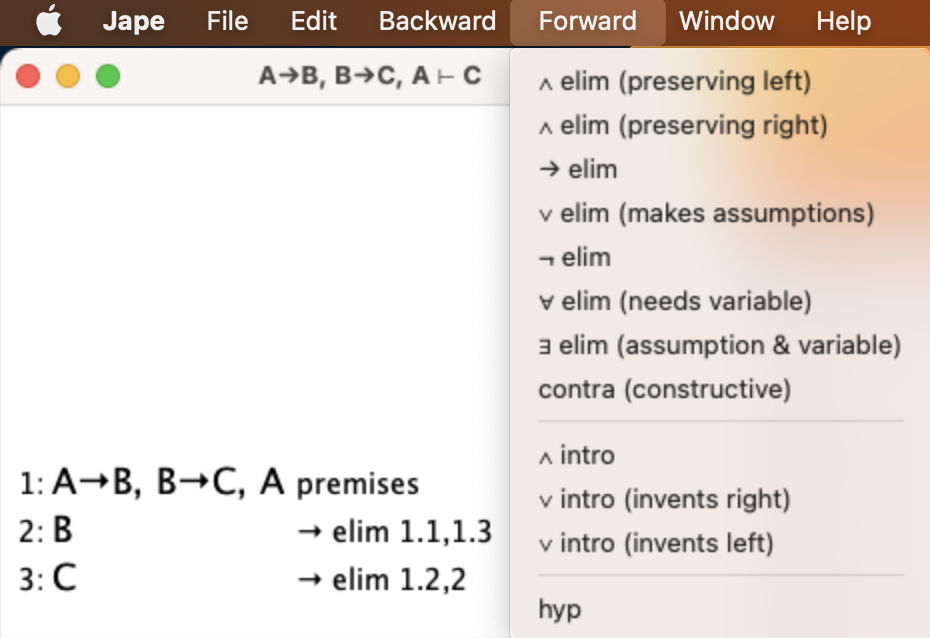}
    \caption{JAPE: $A\rightarrow B, B\rightarrow C, A\vdash C$}
    \label{fig:jape}
\end{subfigure}

\begin{subfigure}{0.5\textwidth}
    \centering
    \includegraphics[scale=0.35]{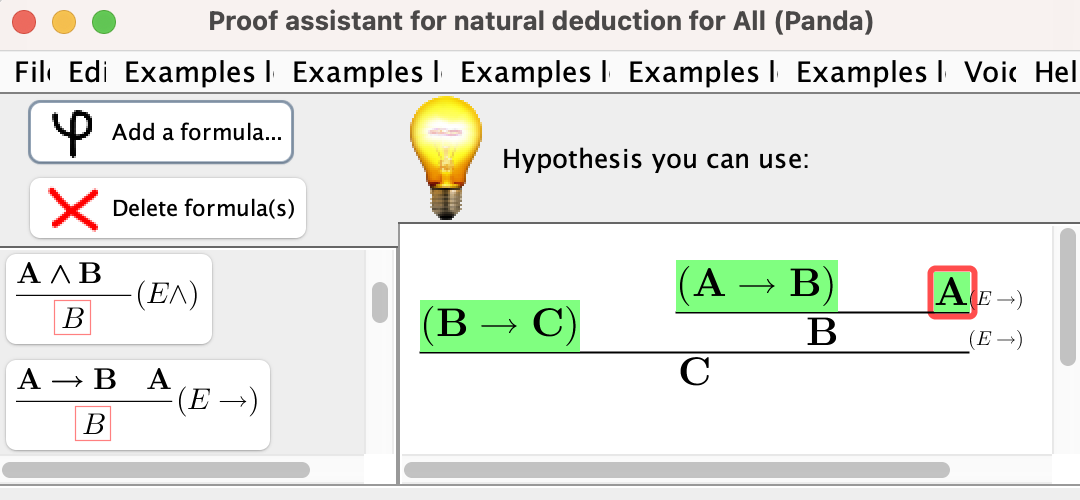}
    \caption{Panda: $A\rightarrow B, B\rightarrow C, A\vdash C$}
    \label{fig:panda}
\end{subfigure}
\begin{subfigure}{0.45\textwidth}
    \centering
    \includegraphics[scale=0.35]{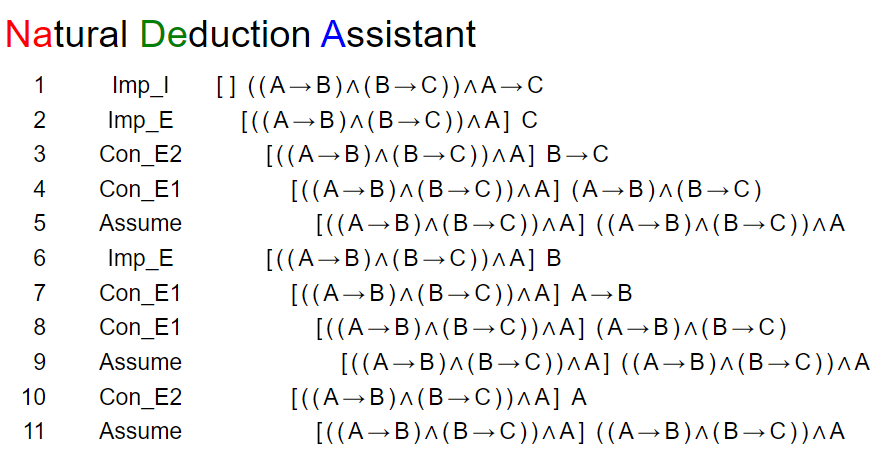}
    \caption{NaDeA: $A\rightarrow B, B\rightarrow C, A\vdash C$}
    \label{fig:nadea}
\end{subfigure}

\caption{Natural Deduction Tableau Proof Assistants}
\end{figure}

\begin{figure}[H]
    \centering
\begin{subfigure}{0.45\textwidth}
    \centering
    \includegraphics[scale=0.28]{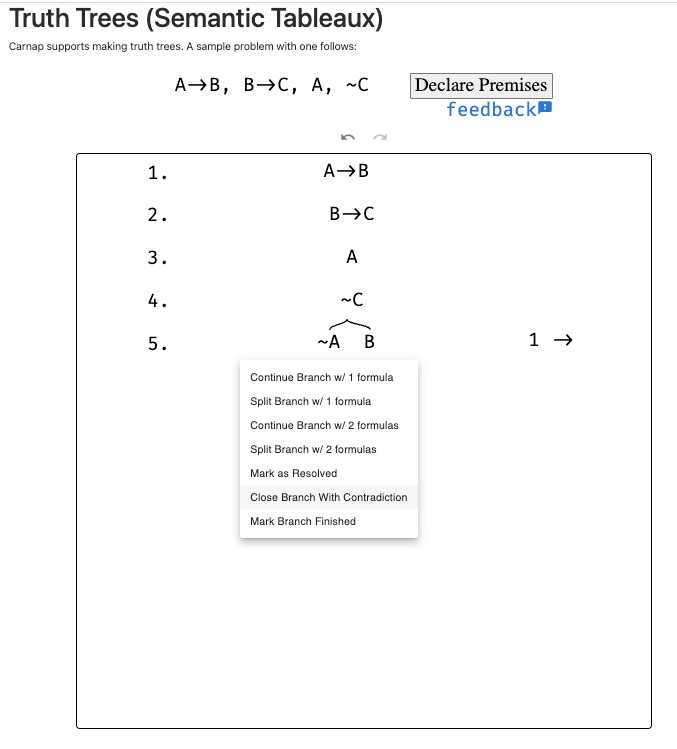}
    \caption{Carnap.io: $A \rightarrow B, B\rightarrow C, A \vdash C$}
    \label{fig:carnap}
\end{subfigure}
\begin{subfigure}{0.45\textwidth}
    \centering
    \includegraphics[scale=0.15]{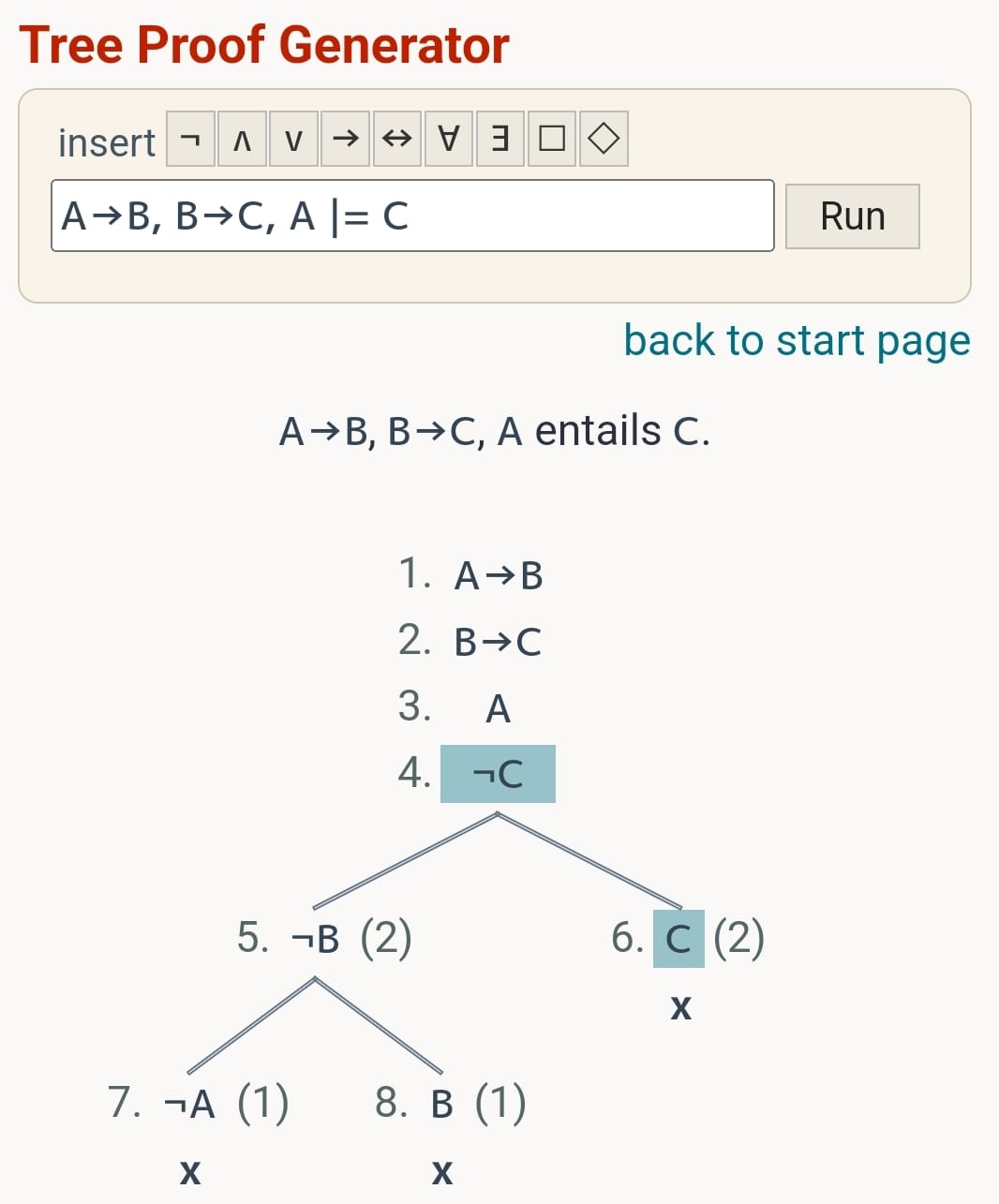}
    \caption{Proof Generator: $A \rightarrow B, B\rightarrow C, A \vdash C$}
    \label{fig:proof-generator}
\end{subfigure}
\caption{Analytic Tableau Proof Assistants}
\end{figure}

Figure~\ref{fig:proofweb} presents a proof in ProofWeb. Note that the student has to learn a new syntax that differs a lot from what the student would write on paper. On the other hand, the proofs in Jape (Figure~\ref{fig:jape}), Panda (Figure~\ref{fig:panda}), NaDeA (Figure~\ref{fig:nadea}), and Carnap.io (Figure~\ref{fig:carnap}) are carried out by the GUI and the user should click on the menu to add each inference rule. The proof generator, in fact, is a prover instead of proof assistant. So, the user can only interact with the tool to enter the theorem to be get either a countermodel or a tree proof that it is not very useful in order to teach how to use the inference rules.

Below we summarize the assistant proofs regarding to: the deductive systems (ND for Natural Deduction, AT for Analytic Tableaux); the display of the proof-style (F for Fitch-style, G for Gentzen-style); The input proof-writing (GUI for based on clicking in the GUI interface, PT for plain text).
\begin{center}
\footnotesize
\noindent\begin{tabular}{|c|c|c|c|c|c|c|c|c|c|}
\hline
 & ProofWeb & Jape & Panda & NaDeA & NADIA & ANITA & Carnap.io & Proof Gen.
\\
\hline
 Deductive Systems & ND & ND & ND & ND & ND & AT & AT & AT 
\\
\hline
Display Proof-Style & F, G & F & G & F & F, G & F & T & T 
\\
\hline
Input Proof-Writing & GUI, PT & GUI & GUI & GUI & PT & PT & GUI & GUI 
\\
\hline
\end{tabular}    
\end{center}

\section{Evaluation of ANITA}
\label{secao:avaliacao}
In this section, we present the results of the evaluation of ANITA that were carried out in two classes of Logic in Computer Science in 2022 at the Federal University of Ceará at Quixadá Campus. Each class has 4 hours of class per week and a total of 16 weeks. The classes had a total of 74 students enrolled.

\subsection{Student Evaluations of ANITA}

\begin{figure}[H]
    \centering
\begin{subfigure}{0.45\textwidth}
    \centering
\begin{tikzpicture}[scale=0.45]
\pie[color={yellow!70, green!70, blue!70, red!70},sum=auto, after number=,text=legend,every only number node/.style={text=black}]{8.3/{1x a week},38.9/{2x a week},50/{3x or more  a week},2.8/{I didn't use it often }}
\end{tikzpicture}
    \caption{How often did you use ANITA?}
    \label{fig:OFTEN-USE-ANITA}
\end{subfigure}
\begin{subfigure}{0.45\textwidth}
    \centering
\begin{tikzpicture}[scale=0.45]
\pie[color={yellow!70, green!70, blue!70, red!70},sum=auto, after number=,text=legend,every only number node/.style={text=black}]{11.1/{ Average},30.6/{ Good},58.3/{ Excellent},0/{Poor}}
\end{tikzpicture} 
    \caption{How do you rate ANITA error messages?}
    \label{fig:UNDERSTAND-ANITA}
\end{subfigure}
\caption{Evaluation}
\end{figure}
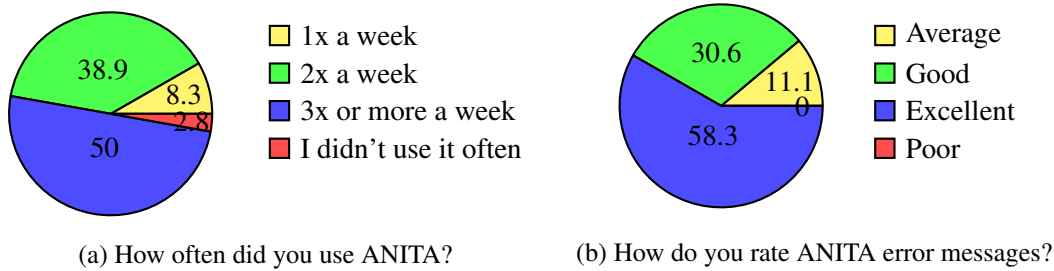

In total 36 out of 74 registered students answered the anonymous online form (49\%). 100\% of the students stated that they used ANITA as a study tool and considered that ANITA helped to exercise the content. 91.7\%  considered ANITA very easy-to-use. Figure~\ref{fig:OFTEN-USE-ANITA} shows how often did the students use ANITA and Figure~\ref{fig:UNDERSTAND-ANITA} 
shows how they rate ANITA error messages. 

\subsection{Evaluation}
\begin{figure}[H]
    \centering
\begin{subfigure}{0.5\textwidth}
    \includegraphics[scale=0.28]{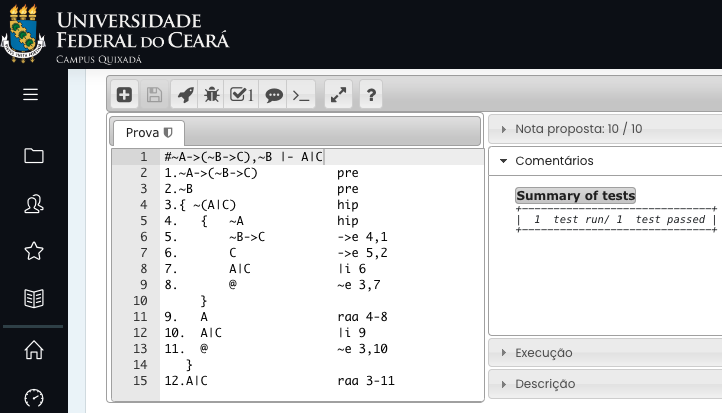}
    \caption{Moodle (NADIA) - Answer of Question 1: \\ $\lnot~A\rightarrow(\lnot~B\rightarrow~C),~\lnot~B\vdash~A\lor~C$}
    \label{fig:Moodle-NADIA}
\end{subfigure}
\begin{subfigure}{0.40\textwidth}
    \includegraphics[scale=0.27]{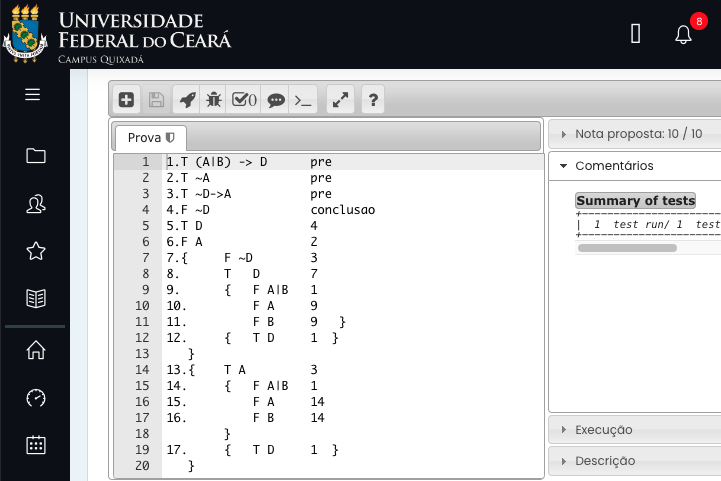}
    \caption{Moodle (ANITA) - Answer of Question 4:\\ $(A\lor B)\rightarrow D, \lnot A, \lnot D \rightarrow A \not\vdash \lnot D$}
    \label{fig:Moodle-ANITA}
\end{subfigure}
\caption{AP2 - Samples of questions of NADIA and ANITA in the Moodle Platform}
\end{figure}

We used NADIA and ANITA, integrated in the Moodle platform\footnote{Moodle is available at \url{https://moodle.com/}}, in the second partial evaluation (AP2), which was applied in the laboratory and had four theorems to be proved in Natural Deduction (ND) and four in analytic Tableaux (AT), each item was worth 1.25. The students wrote down the proofs of each theorem in the Moodle platform and checked automatically, by ANITA and NADIA, whether each proof was correct. For instance, Figure \ref{fig:Moodle-NADIA} displays the answer of a student of Question $1$ of ND in the Moodle platform. Figure \ref{fig:Moodle-ANITA} displays the answer of a student of Question $4$ of AT in the Moodle platform.

In total 48 out of 74 registered students did the test (65\%). The students got a mean (M) of 6.80 with standard deviation (SD) of 3.28. For ND questions, they got 2.76 (MND) and 4.04 (MAT) for AT. 63\% (AND) answered the ND questions, of which 87\% (RND) got the questions right; 92\% (AAT) answered from AT and 88\% (RAT) of those answered the questions correctly. Table \ref{tab:evaluation} presents the results by class.
\begin{table}[H]
\begin{tabular}{|c|c|c|c|c|c|c|c|c|c|c|c|c|c|}
\hline
Class &	Students & M & SD & MND & SDND & AND &	RND & MAT &	SDAT &	AAT & RAT
\\
\hline
A	& 20 & 7.31	& 2.93	& 2.94	& 2.08	& 65\%	& 90\%	& 4.38	& 1.43	& 94\%	& 93\%
\\
\hline
B &	28 & 6.44 & 3.53  & 2.64 & 2.03 & 62\% & 85\% & 3.80 & 1.82 & 91\% & 84\%
\\\hline
A+B & \textbf{48} & \textbf{6.80} & 3.28  & \textbf{2.76} & 2,05 & 63\%	& 87\% & \textbf{4.04} & 1.66 & 92\% & 88\%
\\\hline
\end{tabular}    
    \caption{Results by Class}
    \label{tab:evaluation}
\end{table}

\section{Conclusion and Future Work}\label{sec:conclusao}
ANITA has been used for teaching analytic tableaux to computer science students. We have compared ANITA as a tool for teaching logic to other tools. From the evaluation point of view, ANITA has been a success in our courses. 49\% of the students answered an anonymous online form, in which: 100\% consider that the tool helped to exercise the content; ~91\% consider the tool easy-to-use (excellent or good); ~90\% used the tool two or more times a week; and ~90\% considered the understanding of messages as Excellent or Good. We used ANITA, integrated in the Moodle platform, in the partial evaluation. In total 48 out of 74 registered students did the test (65\%). 92\% of the students submitted their proofs to 4 theorems and of these 88\% got the questions right. 

As future work, we consider developing more teaching materials for ANITA and making further evaluations of ANITA as a tool for teaching logic.

\noindent\textbf{Acknowledgements:} This work is partially supported by the project 04772314/2020/FUNCAP.

\nocite{*}
\bibliographystyle{eptcs}
\bibliography{generic}
\end{document}